\documentclass[11pt]{article}

\usepackage[
    letterpaper,
    margin=1in
]{geometry}

\usepackage[T1]{fontenc}
\usepackage{amsmath,amssymb,amsfonts}
\usepackage{newtxtext,newtxmath}

\usepackage{microtype}

\usepackage{graphicx}
\graphicspath{{media/}}

\usepackage{booktabs}
\usepackage{threeparttable}

\usepackage{caption}
\usepackage{enumitem}
\setlist{nosep}

\usepackage{titlesec}

\titleformat{\section}
    {\large\bfseries}
    {\thesection}
    {0.6em}
    {}

\titleformat{\subsection}
    {\normalsize\bfseries}
    {\thesubsection}
    {0.6em}
    {}

\titleformat{\subsubsection}
    {\normalsize\bfseries}
    {\thesubsubsection}
    {0.6em}
    {}

\titlespacing*{\section}
    {0pt}
    {1.8ex plus .4ex minus .2ex}
    {0.8ex}

\titlespacing*{\subsection}
    {0pt}
    {1.4ex plus .3ex minus .2ex}
    {0.5ex}

\titlespacing*{\subsubsection}
    {0pt}
    {1.2ex plus .3ex minus .2ex}
    {0.4ex}

\makeatletter
\renewcommand{\maketitle}{%
  \thispagestyle{plain}
  \begin{center}
    \vspace*{1em}
    \hrule height 1.2pt
    \vspace{1.1em}

    {\LARGE\bfseries \@title \par}
    \vspace{1.25em}

    {\large
      \def\and{%
        \end{tabular}\hfill\linebreak[0]\hfill%
        \begin{tabular}[t]{c}}
      \begin{tabular}[t]{c}
        \@author
      \end{tabular}\par}
    \vspace{1em}

    {\normalsize \@date \par}
    \vspace{1.1em}

    \hrule height 1.2pt
  \end{center}
  \vspace{1.5em}
}
\makeatother

\usepackage[authoryear,round]{natbib}
\usepackage{fancyhdr}

\newcommand{\shortauthors}{Althaus et al.}
\newcommand{\shorttitle}{Access to Live AI Advice and Behavior Under Risk}

\fancypagestyle{plain}{
    \fancyhf{}
    \fancyfoot[C]{\thepage}
    
}

\usepackage[hidelinks]{hyperref}
\newcommand{\keywords}[1]{%
    \par
    \smallskip
    \noindent
    \textbf{Keywords:} #1
}

\newcommand{\figurenote}[1]{%
    \par
    \medskip
    {\footnotesize #1}
}

\newenvironment{examplebox}
    {\begin{quote}\small\itshape}
    {\end{quote}}

\title{
    Access to Live AI Advice and Behavior Under Risk:
    An Incentivized Experiment
}

\author{
    Paul Althaus \\
    Heidelberg University
    \and
    Leon Houf \\
    Karlsruhe Institute of Technology (KIT) \\
    Leon.Houf@kit.edu
    \and
    Christiane Schwieren
    \\
    Heidelberg University
}

\date{ }

\begin{document}

\maketitle

\begin{abstract}
Generative AI has become an everyday advisor, and the systems people consult are live and interactive, not pre-scripted.
We ask whether access to such a system changes behavior under risk. In an incentivized experiment ($N=158$), participants made lottery choices with an optional decision aid presented as a conventional pre-written tool, a live one-shot AI, or a live interactive AI they could query, with information format held equivalent across conditions. Risk preferences are elicited via DOSE. 
We find no evidence that access to a live AI advisor changes risk aversion.

\keywords{
Risk Aversion; AI advice; Algorithm appreciation; Human–AI interaction; Online experiment
}
\end{abstract}

\section{Introduction}
\label{sec:1}
Generative AI has been adopted at remarkable speed, and the systems people now consult are live and interactive rather than pre-scripted \citep{Bicketal2024, Chatterjietal2025}. Since much of this use seeks advice on consequential choices, a natural question follows: does access to a live AI advisor, rather than a conventional decision aid, change the risks people take?  We test this directly while holding the \textit{informational format} of the advice constant. Participants make incentivized lottery choices with an optional decision aid that is delivered either as a conventional pre-written tool or as a real, live AI advisor, one-shot or interactive. We find no evidence that access to the AI advisor changes the elicited risk aversion (insignificant point estimate close to zero, $\hat \beta_1=0.032$) nor how often the aid is consulted. 

The null result is informative because earlier work gives ample reason to expect the source and delivery of advice to matter. People often place more weight on guidance once it is described as algorithmic \citep{Loggetal2019}, and they can turn against an algorithm after seeing it err \citep{Dietvorstetal2015}, a reaction that itself varies with context and personal traits \citep{ferraz2025trust}. Closer to the choices we study, \cite{ChenFilizOzbayOzbay2026} show that presenting participants with a pre-generated AI recommendation pulls their lottery decisions toward less extreme risk-taking. 

We elicit constant relative risk aversion using the Dynamically Optimized Sequential Experimentation method \citep{Chapmanetal2024}, with payoffs adapted from \cite{HoltLaury2002}, across three arms of a pre-registered experiment\footnote{Pre-registration: \url{https://aspredicted.org/p8qg3i.pdf}. IRB: \url{https://gfew.de/ethik/N1zJ5Lar}} ($N = 158$) that differ only in how the optional aid is presented. Two features make the null informative rather than merely inconclusive. A randomized order effect of comparable magnitude is precisely estimated in the same data, confirming that the design detects effects of the size at issue, and the null persists both among the participants who actually consult the aid and among those who report high trust in AI, so it is not an artifact of non-use. We can rule out effects on risk aversion larger than $\Delta r = 0.275$, while smaller effect sizes remain possible. We are explicit about scope: we identify the effect of access to a live AI advisor net of an equivalent informational format, and we do not claim that the information AI conveys is itself without effect.

\section{Experimental Design}
\subsection{Setup}\label{sec:2.1}
We hold the format of advice equivalent across arms and vary only which decision aid participants have access to.
The experiment was conducted online with participants recruited from the AWI laboratory at Heidelberg University in June 2026 and implemented in oTree \citep{Chenetal2016}. Participants took part on their own devices, mirroring how AI chatbots are typically used. Each participant answered six incentivized lottery questions, choosing between a safe and a risky option; the side on which the safe option appeared was randomized per participant (indicator \textit{Safe = Option A}). The form follows \citet{KahnemanTversky1979} and the payoffs are adapted from \citet{HoltLaury2002}. The first question, shown below, was identical for all participants.
\begin{examplebox}
\begin{center}
\begin{tabular}{p{4cm} p{4cm}}
    \textbf{Option A (Safe):} & \textbf{Option B (Risky):} \\
    38\% chance to win 2.50€ & 38\% chance to win 5.50€ \\
    62\% chance to win 2.00€ & 62\% chance to win 0.10€ \\
\end{tabular}
\end{center}
\end{examplebox}
We use the Dynamically Optimized Sequential Experimentation (DOSE) method rather than a multiple price list, as it estimates $r$ more precisely \citep{Chapmanetal2024}: the first question is fixed and the remaining five adapt to the participant's prior answers. From these choices we estimate an individual Constant Relative Risk Aversion (CRRA) parameter $r$. We recruited 158 participants, and paid them a fixed fee of 2.50€ plus the payoff of one randomly selected round.

\subsection{Treatment Arms}\label{sec:2.2}
Participants were randomly assigned to one of three arms (Control: 56, One-Shot AI: 53, Interactive AI: 49). In every arm an optional decision aid is available by clicking on a button. The control arm provides the information alone: a pre-programmed aid in oTree that automatically calculates and reports both options' expected values and notes what a risk-averse or risk-tolerant person might do. The one-shot arm presents the same information through a live AI (``AI Assistance''), consulted once per round. The interactive arm (``Interactive AI Assistance'') adds the possibility to query it, with up to two follow-up questions per round. 

To keep the informational format as equivalent as possible with a live system, the AI (gpt-5.4-mini by OpenAI) was system-prompted to report each option's expected value and to use the same safe/risky framing as the pre-programmed aid. It computed the safe option's expected value correctly in every consultation; for the risky option it was correct in 216 of 228 (94.7\%), erring in 12. While we consider AI errors as part of the technology and therefore the treatment, we also perform subgroup analysis to exclude these cases. After the lotteries, participants completed a short questionnaire on demographics and attitudes toward AI, including the Short Trust in AI Scale (S-TIAS) \citep{McGrathetal2025}.

\section{Analysis Strategy}\label{sec:analysis}
Our dependent variable is the DOSE posterior mean of each participant's risk-aversion parameter, $\hat r_i$. We estimate
\begin{equation}\label{eq:main}
\hat r_i = \beta_0 + \beta_1\,\text{OneShot}_i + \beta_2\,\text{Interactive}_i + \varepsilon_i,
\end{equation}
where $\beta_1$ is the pre-registered, confirmatory contrast of one-shot AI against control (H1) and $\beta_2 - \beta_1$ is the pre-registered, exploratory contrast of interactive against one-shot AI (H2). Standard errors are HC3 robust, and we assess robustness to actual tool use, S-TIAS trust, the decision-noise parameter $\lambda$, and the DOSE posterior uncertainty in $\hat r$.

\section{Results}\label{sec:results}
\subsection{Does AI access shift risk aversion? }
\begin{figure}[t]
\centering
\caption{Histogram of Estimated $r$ per Arm}
\label{fig:r_Histogram_Arms}
\includegraphics[width=1\linewidth]{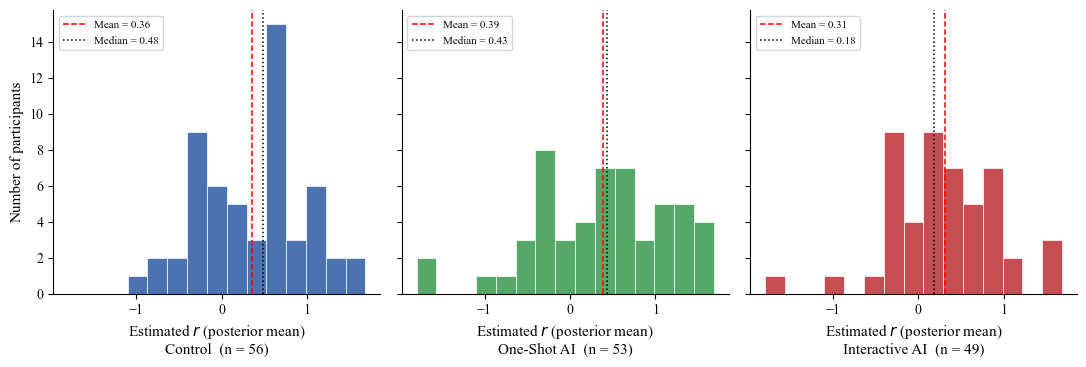}
\figurenote{\underline{Note:} All three x-axes show the CRRA $r$, which is defined on the interval [-2, 2]. The y-axis shows the  respective number of participants.}
\end{figure}

We do not find any significant effects of access to AI on the estimated CRRA parameter $\hat r$. 
Table~\ref{tab:full} reports estimates of Equation~\eqref{eq:main}. Access to a one-shot AI leaves risk aversion essentially unchanged: $\hat\beta_1 = 0.032$ (95\% CI $-0.239,\,0.302$), a point estimate close to zero. Access to an interactive AI adds nothing beyond this, as the contrast between the interactive and one-shot arms is $\hat\beta_2 - \hat\beta_1 = -0.077$ ($F = 0.28$, $p = 0.60$). Only the intercept is significant, implying moderate risk aversion with \(\hat{r} = 0.356\), which matches what other experimental studies find \citep{HoltLaury2002, EbertWiesen2014, trautmann2018higher}. The only coefficient to differ from zero is the randomized position of the safe option. The distribution of $\hat r$ is visibly similar across arms (Figure~\ref{fig:r_Histogram_Arms}). 

\begin{table*}[t]\centering
\begin{threeparttable}
  \caption{Primary and Robustness Analysis: Effect of AI Access on Estimated $r$}
  \label{tab:full}
\setlength{\tabcolsep}{4pt}
\begin{tabular}{lccccc}
\toprule
& \multicolumn{1}{c}{Primary} & \multicolumn{1}{c}{+ gender} & \multicolumn{1}{c}{+ course} & \multicolumn{1}{c}{+ position} & \multicolumn{1}{c}{Full} \\
\midrule
One-Shot AI (vs.\ Control)    & 0.032   & 0.039   & 0.030   & 0.040   & 0.040   \\
                              & (0.138) & (0.140) & (0.139) & (0.137) & (0.140) \\
Interactive AI (vs.\ Control) & -0.045  & -0.044  & -0.048  & -0.030  & -0.037  \\
                              & (0.122) & (0.125) & (0.123) & (0.118) & (0.120) \\
Gender: Male                  &         & -0.103  &         &         & -0.085  \\
                              &         & (0.111) &         &         & (0.109) \\
Gender: Prefer not to tell    &         & -0.059  &         &         & 0.146   \\
                              &         & (0.852) &         &         & (0.853) \\
Took relevant courses         &         &         & 0.075   &         & 0.100   \\
                              &         &         & (0.110) &         & (0.107) \\
Safe = Option A               &         &         &         & -0.308$^{**}$ & -0.308$^{**}$ \\
                              &         &         &         & (0.108) & (0.112) \\
Intercept                     & 0.356$^{***}$ & 0.400$^{***}$ & 0.325$^{***}$ & 0.504$^{***}$ & 0.500$^{***}$ \\
                              & (0.080) & (0.091) & (0.096) & (0.101) & (0.119) \\
\midrule
Observations        & 158   & 158   & 158   & 158   & 158   \\
\bottomrule
\addlinespace[0.3ex]

\end{tabular}
\begin{tablenotes}[flushleft]
\footnotesize
\item $^{*}$p$<$0.05; $^{**}$p$<$0.01; $^{***}$p$<$0.001.
HC3 robust SEs in parentheses. $N=158$.
Treatment coefficients are stable across specifications.
\textbf{Course:} Has taken relevant economic courses.
\textbf{Position:} Safe Option is Option A.
Only the (randomized) position control is significant.
\end{tablenotes}
\end{threeparttable}
\end{table*}

The null is descriptively stable across relevant subsamples. Treatment estimates remain close to zero and insignificant among participants who consulted the aid at least once, twice, or three times, and among those with above-median trust in AI (Table~\ref{tab:subgroups}). While these post-treatment comparisons are not causal, they provide no indication that the aggregate null is concentrated among non-users or participants with lower AI trust. The result survives dropping the participants exposed to a misstated risky EV by the AI. With these 11 participants excluded, the treatment effects remain small and insignificant (one-shot $+0.05$, interactive $-0.07$, $p>0.5$).

\begin{table*}[t]\centering
\begin{threeparttable}

  \caption{Subgroup Robustness: Does the Null Persist?}
  \label{tab:subgroups}
\setlength{\tabcolsep}{5pt}
\begin{tabular}{lccccc}
\toprule
& \multicolumn{1}{c}{Used $\geq$1} & \multicolumn{1}{c}{Used $\geq$2} & \multicolumn{1}{c}{Used $\geq$3} & \multicolumn{1}{c}{High trust} & \multicolumn{1}{c}{Excl.\ EV error}\\
\midrule
One-Shot AI (vs.\ Control)    & -0.014  & 0.019   & -0.088  & 0.017  &  0.051 \\
                              & (0.166) & (0.180) & (0.198) & (0.180) & (0.143) \\
Interactive AI (vs.\ Control) & -0.011  & 0.112   & 0.092   & -0.109  & -0.074\\
                              & (0.145) & (0.132) & (0.140) & (0.185)  & (0.134)\\
Intercept                     & 0.327$^{**}$ & 0.241$^{*}$ & 0.205$^{}$ & 0.354$^{***}$ &  0.356$^{***}$\\
                              & (0.103) & (0.100) & (0.107) & (0.103) & (0.080) \\
\midrule
Observations        & 93    & 71    & 59    & 83  & 147  \\
\bottomrule

\end{tabular}
\begin{tablenotes}[flushleft]
\footnotesize
\item $^{*}$p$<$0.05; $^{**}$p$<$0.01; $^{***}$p$<$0.001.
HC3 robust SEs in parentheses.
Cols.\ 1--3: tool used in $\geq k$ rounds.
Col.\ 4: above-median AI trust.
Col.\ 5: participants who received a misstated risky EV
($>$€$0.02$) excluded.
\end{tablenotes}

\end{threeparttable}
\end{table*}

As pre-registered, we re-ran the primary regression with the upper and lower bound of the 95\% Credible Interval for each participant and with alternative values of 0.3 and 0.7 for the decision noise parameter $\lambda$. All treatment estimates are close to zero and non-significant.

\subsection{Is the null informative?}
A null is informative only if the design could have detected an economically meaningful effect, and two things establish that it could even while the realized between-subject SD of $\hat r$ was $0.68$ against the $0.40$ assumed at pre-registration. First, the design is demonstrably sensitive: the randomized side of the safe option shifts $\hat r$ by $-0.308$ ($p<0.01$), an effect orthogonal to treatment by construction and larger than the one we set out to detect, so the design registers differences of the relevant size when they are present. Second, we bound the treatment effect directly. Our smallest effect of interest is the pre-registered $\Delta r = 0.275$. Two one-sided tests (TOST) reject effects beyond this bound for both treatment arms against control ($p = 0.040$ for one-shot, $0.031$ for interactive, 90\% CI), and Bayes factors of $4.8$ and $4.5$ (Cauchy prior, scale $0.71$) favor the null over the alternative. Both point estimates sit near zero, and the intervals stay within the bound, so we can rule out effects large enough to matter, while smaller effects remain possible; the exploratory one-shot-versus-interactive contrast is underpowered for either verdict as pre-registered.

\subsection{Does the medium change usage?}
Consultation of the aid does not differ significantly across arms, though it trends upward with the richness of the aid, from 50.0\% under the pre-programmed aid to 67.3\% in the interactive arm (Table~\ref{tab:usage}; control vs.\ pooled AI: $\chi^2=2.27$, $p=0.132$). The trend is exploratory and not pre-registered. Even so, the interactive channel itself went almost unused, with only 2 of 49 participants ever asking a follow-up, so any pull of the AI advisor on usage operates through the initial consultation, not sustained engagement.

\begin{table}[t]
\centering
\caption{Aid consultation by treatment arm}
\label{tab:usage}
\begin{threeparttable}
\begin{tabular}{lccc}
\toprule
 & Consulted & Never & Rate \\
\midrule
Control (pre-programmed) & 28 & 28 & 50.0\% \\
One-Shot AI              & 32 & 21 & 60.4\% \\
Interactive AI           & 33 & 16 & 67.3\% \\
\bottomrule
\end{tabular}
\begin{tablenotes}[flushleft]
\footnotesize
\item \underline{Note:} $N = 158$ (56, 53, 49). ``Consulted'' counts participants who opened the aid at least once. Control vs.\ pooled AI arms: $\chi^2(1) = 2.27$, $p = 0.132$. Across arms: $\chi^2(2) = 3.32$, $p = 0.190$. Exploratory, not pre-registered.
\end{tablenotes}
\end{threeparttable}
\end{table}

\section{Discussion}\label{sec:discussion}
Offering a live AI advisor in place of a conventional aid that carried equivalent information, we find no effect on $\hat r$, i.e., participants were no more or less risk averse, and they consulted the aid not significantly more often. 

Prior work gave reason to expect a shift. Advice can gain weight simply because it is labeled algorithmic \citep{Loggetal2019, jussupow2024integrative}, and, closest to our task, an imposed AI recommendation has been shown to move lottery choices \citep{ChenFilizOzbayOzbay2026}. We read the null result in our experiment as a sign that this label effect is not fixed but context-dependent. It may depend on how established the technology is, and AI advice is by now common enough that the effect may have faded. Reactions to algorithms soften as people gain experience with them \citep{filiz2021reducing, dietvorst2018overcoming}, and our participants are drawn from the young, frequent-use group for whom such advice is already routine \citep{Bicketal2024}, reporting trust that is moderate rather than enthusiastic. One possibility for an explanation of the null result might be that whatever novelty a live advisor once carried may simply have worn off. This yields several predictions: the medium might matter more where AI is less familiar, or where advice is framed as always present recommendation; higher stakes may restore it as well. Other explanations remain open, from the reasoning limits of current models \citep{Shojaeeetal2025} to a general wariness of algorithmic advice \citep{Opreaetal2024}.

One feature of studying a live AI advisor deserves comment. We held the basic form of the advice, the expected values, and the safe/risky framing constant via the system prompt, but the live AI model occasionally computed an expected value wrong, which the pre-programmed aid never would. We read this as part of the treatment rather than a flaw: a real AI advisor is fallible, and that fallibility is a property of the medium we set out to study, not something to be controlled away. In our data the errors were rare, arising in about $5.2\%$ of consultations, and show no sign of having moved behavior, the risky option they inflated was already the higher-EV choice, and the participants who saw an inflated value were, if anything, slightly more risk-averse than others in their arm. What we cannot do is estimate the errors' influence; they are too infrequent for that, and we take no position on how AI mistakes would matter if they were common.

Three limitations qualify the result. First, we study \textit{access} to an advisor, not a requirement to use it, neither in the one-shot AI nor the interactive AI; it has been shown that an imposed recommendation can move behavior \citep{ChenFilizOzbayOzbay2026}, so the null speaks to the effect of availability only. Second, we hold one message regarding economic expected value equivalent across arms. A different message, or a model that behaved differently, might well move behavior, and a live system's output can never be fully controlled; the same prompt does not need return the same answer twice. That is a limitation, but it is also what makes this a test of the current-day technology people use, on modest stakes and a single population. None of it undoes the null. The effect stays near zero among participants who consulted the aid and among those who trust AI most, and the design still registers a randomized display effect of comparable size. Third, due to the larger standard deviation than expected, the study results can rule out effect sizes greater than $\Delta r = 0.275$. Smaller effect sizes remain possible, while the point estimate of the current data is very close to zero ($\hat \beta_1 = 0.032$).

Access to a live AI advisor, then, does not by itself change how people take risks in an economically meaningful manner. This suggests where effects may be in a population that has grown accustomed to live AI advisors: in whether advice is consumed and imposed rather than merely available. The interaction of familiarity with a medium and its effect through advice on behaviour is a question for further research both for AI advisors and for any other new technology offering advice.

\section*{Acknowledgments}

The study was funded by Heidelberg University. Leon Houf acknowledges funding by the Joachim Hertz Stiftung and by the Deutsche Forschungsgemeinschaft (DFG, German Research Foundation) -- GRK2739/2 -- Project Nr.\ 447089431 -- Research Training Group KD\textsuperscript{2}School: Designing Biosignal-Adaptive Systems for Decision-Making Processes. 

\section*{Data and Code}

All instructions, system-prompt, full results and data and code will be made available in the supplementary material upon publication and are available now by request.

\bibliographystyle{plainnat}
\bibliography{references}

\end{document}